# Dynamics in direct two-photon transition by frequency combs

Lin Dan, Hao Xu, Ping Guo and Jianye Zhao[*]

*Department of Electronics, Peking University, Beijing 100871, People's Republic of China*

Two-photon resonance transition technology has been proven to have a wide range of applications, it's limited by the available wavelength of commercial lasers. The application of optical comb technology with direct two-photon transition (DTPT) will not be restricted by cw lasers. This article will further theoretically analyze the dynamics effects of the DTPT process driven by optical frequency combs. In a three-level atomic system, the population of particles and the amount of momentum transfer on atoms are increased compared to that of the DTPT-free process. The 17% of population increasement in 6-level system of cesium atoms has verified that DTPT process has a robust enhancement on the effect of momentum transfer. It can be used to excite the DTPTs of rubidium and cesium simultaneously with the same mode-locked laser. And this technology has potential applications in cooling different atoms to obtain polar cold molecules, as well as high-precision spectroscopy measurement.

## I. INTRODUCTION

Interaction of atoms with continuous dichromatic laser can give rise to two-photon resonances based on cascade three energy level system. This technology has been used for two-photon Doppler cooling of alkaline-earth-metal [1], as well as metastable helium atoms [2], hydrogen atoms [3] and so on. This method is limited by the availability of cw lasers to a subset of atoms and molecules that have complex internal structure. With the development and wide applications of femtosecond optical frequency combs (FCs) [4-6], many scientific researches like fundamental physics measurements [7], ultrafast optics, strong optical field, ultraviolet and deep ultraviolet measurement [8] have been revolutionized and achieved great advances. The application of FCs together with Two-photon transitions (TPTs) generates a new subject named direct frequency comb spectroscopy [9], in addition, high-precision spectroscopy measurement [10,11] and the optical frequency standard [12] can also be realized. As a new tool for manipulating atoms, Kielpinski proposed a scheme of using direct two-photon transition (DTPT) to laser cool atoms and molecules [13], and Jayich *et al*. observed the cooling phenomenon in experiments by using this technology [14].

In this paper, we are going to further theoretically investigate the dynamics effects of DTPT induced by frequency combs and compare them with the results of DTPT-free situation. The paper will be organized in the following manner: Sec. II will describe the effects of DTPT based on the three-level atomic system, including the characteristics of the two-photon Rabi frequency, the difference between the population of particles in the upper energy level under the Doppler effect and that without DTPT, and the change in momentum transfer produced by DTPT. In Sec. III, we will extend the three-level results to a multi-level system where DTPT may occur and the transition wavelengths are close to each other. Taking the alkali metal cesium atom as an example, we get the optical Bloch equations (OBEs) of the 6-level system, and obtain population difference that are relative to the DTPT-free case. The population change has verified that DTPT has a better effect on the transfer of atom-photon momentum. Sec. IV will discuss the application of the theoretical analysis in the previous article in practice. By comparing the existing observation results, we find that the same mode-locked laser can simultaneously excite the DTPTs of rubidium and cesium. And this technology of manipulating multiple elements at the same time provides a wide range of applications. Finally, Sec. V presents the conclusion.

---

* zhaojianye@pku.edu.cn.

## II. DYNAMICS IN DTPT

Considering a three-energy level atomic system interacting with FCs, it includes two TPTs processes, one is the resonant TPTs process, atoms absorb photons with angular frequencies $\omega_1$ going from ground state 1 to intermediate state 2, and then absorb photon with angular frequencies $\omega_2$ going from state 2 to final state 3, and the other non-resonant TPT process going from ground state 1 to exited state 3 by absorbing two photons with sum frequencies equal $\omega_3$ directly. The population dynamics of these three states can be described by density-matrix elements $\rho_{ij}$ (i,j=1,2,3), which can be calculated by solving the OBEs under the rotating-wave approximation given by

$$\dot{\rho}_{33} = i(\Omega_{13}^*\sigma_{13}+\Omega_{23}^*\sigma_{23} - c.c.) - \Gamma_{33}\rho_{33}$$
$$\dot{\rho}_{22} = i(\Omega_{12}^*\sigma_{12}-\Omega_{23}^*\sigma_{23} - c.c.) + \Gamma_{33}\rho_{33} - \Gamma_{22}\rho_{22}$$
$$\dot{\sigma}_{23} = i[\delta_{23}\sigma_{23}+(\Omega_{12}^*\sigma_{13} - c.c.) + \Omega_{23}(\rho_{33} - \rho_{22})]$$
$$- \Gamma_{23}\sigma_{23}$$
$$\dot{\sigma}_{12} = i[\delta_{12}\sigma_{12} + \Omega_{12}(2\rho_{22} + \rho_{33} - 1) + \Omega_{13}\sigma_{23}^*$$
$$- \Omega_{23}^*\sigma_{13}] - \Gamma_{12}\sigma_{12}$$
$$\dot{\sigma}_{13} = i[\delta_{13}\sigma_{13} + \Omega_{13}(2\rho_{33} + \rho_{22} - 1) + \Omega_{12}\sigma_{23} -$$
$$\Omega_{23}\sigma_{12}] - \Gamma_{13}\sigma_{13}$$

(1)

Where the frequency detuning is defined as $\delta_{12} = \omega_{12} - \omega_1$, $\delta_{23} = \omega_{23} - \omega_2$, $\delta_{13} = \omega_{13} - \omega_3$, $\omega_1$ and $\omega_2$ are single photon frequency that drive transition from 1 to 2 and 2 to 3 respectively, $\omega_3=\omega_1+\omega_2$ is the sum frequency of two photons that drive the transition from 1 to 3 directly. The frequencies between two energy levels are defined as $\omega_{ij} = (E_j - E_i)/\hbar$. $\Gamma_{ij}$ is the spontaneous decay rate, $\sigma_{12}$, $\sigma_{12}$, $\sigma_{12}$ are the transformed off-diagonal elements of the density matrix and are related by $\sigma_{12} = \rho_{12}e^{-i\omega_1 t}$, $\sigma_{23} = \rho_{23}e^{-i\omega_2 t}$, $\sigma_{13} = \rho_{13}e^{-i(\omega_1+\omega_2)t}$.

### A. Two-photon Rabi frequency

In order to obtain the dynamic characteristics of excited state population from OBEs, we extract available parameters from common alkali metal atom transitions, such as Rabi frequency and excited state spontaneous emission rate. We define the single photon Rabi frequency as $\Omega_{ij} = \mu_{ij}\varepsilon/\hbar$ and the two-photon Rabi frequency [15] as

$$\Omega_{if} = \beta_{if}I \quad (2)$$

where $\mu_{ij}$ is the dipole moment of the corresponding transition, $\varepsilon = E(t)e^{-i\omega_i t}$ is the slowly varying envelope of the laser pulse, $I$ is the intensity of the laser beam, the coefficient [15] is

$$\beta_{if} = -\frac{e^2}{c\varepsilon_0\hbar^2}\Sigma_r\left[\frac{\langle f|x|r\rangle\langle r|x|i\rangle}{\omega_1-\omega_{ir}-\frac{i\Gamma_r}{2}} + \frac{\langle f|x|r\rangle\langle r|x|i\rangle}{\omega_2-\omega_{ir}-\frac{i\Gamma_r}{2}}\right]. \quad (3)$$

The two-photon Rabi frequency is depending on both the laser intensity and the laser frequency detuning, for a train of pulses travelling in x direction with intensity of 0.65Wm$^{-2}$ and beam waist width at about 1.5mm, the two-photon Rabi frequency for an alkaline metal atom vary with both two-photon detuning $\Delta$ and single photon detuning $\delta$ as shown in Fig.1(a), in which $\Delta= \omega_1 + \omega_2 - \omega_{if}$, and $\delta = \omega_1 - \omega_{ir}$, i, r, f indicate initial ground state, intermediate state, final state respectively. Although the corresponding peak values of different alkali metals are different, the Rabi frequency as shown in Fig.1 can always be obtained by adjusting the light intensity. The peak value actually represents the value of the two-photon resonance transition enhancement, that is, the value when $\Delta= \delta=0$. For non-resonant of the two-photon transition, we can take the rabi frequency as the half-peak value for the calculation in the following text, that is about 0.5 MHz as shown in Fig.1(b). And although theoretically the Rabi frequency changes continuously with the single-photon detuning as shown in Fig.1(b), for the optical comb, when the laser detuning frequency becomes larger, the probability of the two-photon transition is almost negligible, which situation will be seen in Sec. IV.

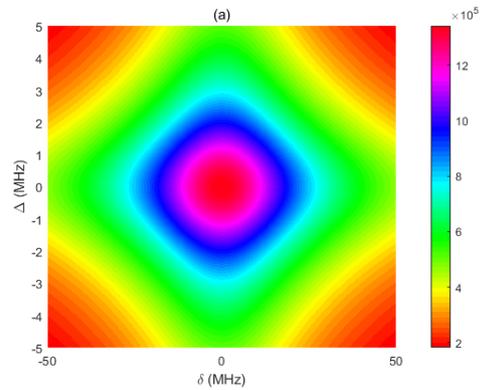

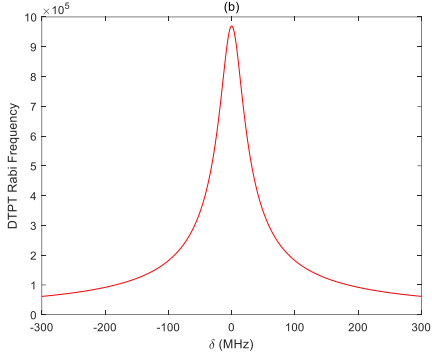

FIG. 1. DTPT Rabi frequency vary with (a) $\delta$ and $\Delta$, (b) $\delta$ only, the half maximum is about 0.5MHz.

### B. Population distribution

To clearly illustrate the effects in the DTPT process, we compare the results obtained from Eqs. (1) by employing the method introduced by Felinto *et at*. [16]. Assuming atoms are initially at the ground state and interact with a train of Gaussian-shaped pulses, the envelope function is $E(t) = E_0 \exp(-\frac{t^2}{2T_p^2})$. We set the pulse width $T_p = 50 fs$ and the pulse repetition period equals 10 ns, the typical values of Rabi frequencies and decay rates for alkali atoms are as following: $\Omega_{12} = 0.70 MHz$, $\Omega_{23} = 0.16 MHz$, $\Omega_{13} = 0.50 MHz$, $\Gamma_{33} = 4.1 MHz$, $\Gamma_{22} = 37 MHz$, $\Gamma_{12} = 18 MHz$, $\Gamma_{23} = 21 MHz$, $\Gamma_{13} = 2.1 MHz$.

The population distribution of the upper state $\rho_{33}$ depend on atomic velocity is shown in Fig. 2. For the Doppler frequency shift caused by atomic motion, the population of particles at the upper energy level is related to the amount of frequency shift. Different velocities of the atoms are corresponding to different frequency detuning. The red solid line indicates that the population of particles is related with the changing of velocity when DTPT is considered. The blue dashed line indicates the situation where DTPT is not occurred. The peak of the figure shows the constructive interference. This is because the pulse interval is less than the life of the upper energy level state, which leads to the accumulation of the population of coherent particles. Considering the peak line for DTPT in contrast to that with no DTPT, the increment in amplitude and width can be regarded as the result of the increase in optical power.

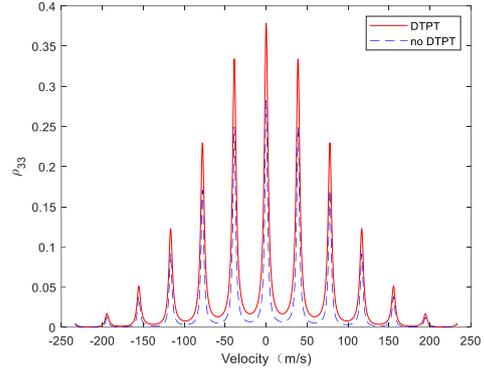

FIG. 2. Uper state population distribution vary with atomic velocity. The red solid line represents the particle population distribution produced by the DTPT process, and the blue dashed line represents the result of the DTPT-free process

### C. Momentum transfer

The momentum transferred on a three-level atom subjected to TPT process can be calculated by [17]

$$\Delta P = \sum_i \hbar k_i \rho_{ii} \qquad (4)$$

Where $\hbar k_i$ is the momentum transfer by one pulse when atoms are excited from ground state to excited state. As shown in Fig.3, the momentum kick reaches maximum at specific velocity, which compensates for Doppler shift caused by atomic motion. Red (blue) line indicates the atom-pulse interaction with (without) DTPT, it's clearly shown that the momentum transfer on atoms increase if we consider the DTPT process.

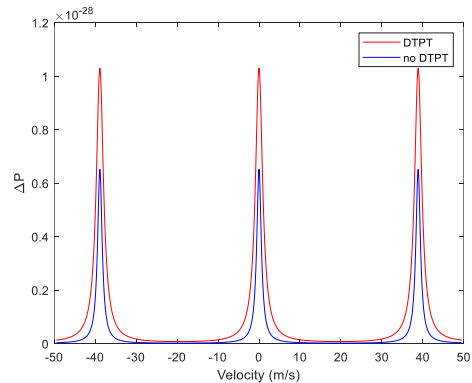

FIG. 3. Momentum transfer at different velocity due to one pulse, the red and blue lines indicate DTPT and DTPT-free respectively.

For more pulses interaction, we have compared our results with the single photon transition. Previous studies have shown that TPT process has a lower Doppler cooling temperature [1], from Fig. 4 we see that the momentum transfer on atoms driven by single photon transition is slightly smaller than that driven by traditional TPT, and four times smaller than that with DTPT process.

Here we also investigate the pulse shape that would affect the corresponding results, we choose a train of hyperbolic-secant pulses, the red line as shown in Fig. 4, all the other parameters are the same as Gaussian-shaped pulses, the train of pulses interact with atoms initially at its ground state, considering the DTPT process, the momentum transfer to atoms tends to not increase after 170 pulses of interaction, while it needs a number of 180 Gaussian pulses to reach a steady state. The pulse numbers and the maximum change of momentum are depending on the TPT Rabi frequency. For the same two-level and three-level atom system, without considering the DTPT process, only about 20 and 100 pulses are needed respectively to obtain the equilibrium state.

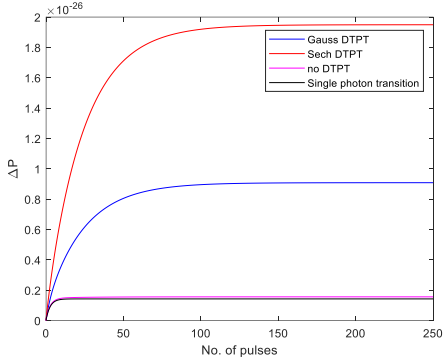

FIG. 4. Momentum transfer varies with the number of pulses, the red line indicates the hyperbolic-secant pulses that contain the result of the DTPT effect, the others are Gaussian pulses, the blue line includes the DTPT process, while the purple line doesn't, and the black line indicates the result of a single photon transition.

### III. MULTILEVEL DTPT APPLICATION

For a practical application of DTPT, we consider the fine structure of isotope $^{133}$Cs as shown in Fig.5, six levels with transition wavelengths close to each other are illustrated, the corresponding transition wavelengths from 761nm-921nm are also labeled.

As discussed above, in the three-level system, the atom obtains a greater momentum transfer after considering the direct two-photon process. Previous study has observed the pushing force on atoms during TPT process in the $^{87}$Rb experiment [14] too, and now it is extended to six-energy level system to further studies its effect on atoms, which is applicable to alkali metal atoms, because the photon transition frequencies between these energy levels are close to each other, such as atomic energy level structure of cesium as shown in Fig.5, the wavelength range marked in the figure is concentrated at 761-921nm. Mode-locked lasers can fully meet the requirements of these transition frequencies at the same time. We will compare these transitions of Cs with Rb in details later.

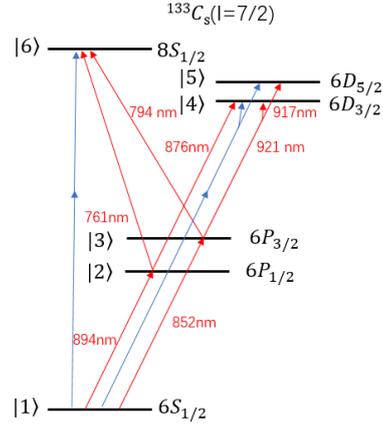

FIG. 5. The 6 energy level structure corresponding to the similar transition wavelength of cesium atom.

We describe the OBEs in Eqs. (5) for the alkali metal 6-level system, all the parameters are defined the similar way as we do for the 3-level system. Through numerical solution, we obtain the population change of atoms as shown in Fig. 6, the momentum transfer is related to the particle population distribution according to Eq. (4).

$$\dot{\rho}_{66} = i(\Omega_{61}\sigma_{16} + \Omega_{62}\sigma_{26} + \Omega_{63}\sigma_{36} - c.c.) - \rho_{66}\Gamma_{66}$$
$$\dot{\rho}_{55} = i(\Omega_{51}\sigma_{15} + \Omega_{53}\sigma_{35} - c.c.) - \rho_{55}\Gamma_{55}$$
$$\dot{\rho}_{44} = i(\Omega_{41}\sigma_{14} + \Omega_{42}\sigma_{24} + \Omega_{43}\sigma_{34} - c.c.) - \rho_{44}\Gamma_{44}$$
$$\dot{\rho}_{33} = i(\Omega_{31}\sigma_{13} + \Omega_{34}\sigma_{43} + \Omega_{35}\sigma_{53} + \Omega_{36}\sigma_{63} - c.c.) - \rho_{33}\Gamma_{33} + \rho_{44}\Gamma_{44}/2 + \rho_{55}\Gamma_{55}/2 + \rho_{66}\Gamma_{66}/2$$
$$\dot{\rho}_{22} = i(\Omega_{21}\sigma_{12} + \Omega_{24}\sigma_{42} + \Omega_{26}\sigma_{62} - c.c.) - \rho_{22}\Gamma_{22} + \rho_{44}\Gamma_{44}/2 + \rho_{66}\Gamma_{66}/2$$
$$\dot{\sigma}_{12} = i(\Omega_{12}\rho_{22} + \Omega_{14}\sigma_{42} + \Omega_{16}\sigma_{62} + \delta_{12}\sigma_{12} -$$

$$\Omega_{12}\rho_{11} - \Omega_{42}\sigma_{14} - \Omega_{62}\sigma_{16}) - \sigma_{12}\Gamma_{12}$$
$$\dot{\sigma}_{13} = i(\Omega_{13}\rho_{33} + \Omega_{14}\sigma_{43} + \Omega_{15}\sigma_{53} + \Omega_{16}\sigma_{63} + \delta_{13}\sigma_{13} - \Omega_{13}\rho_{11} - \Omega_{43}\sigma_{14} - \Omega_{53}\sigma_{15} - \Omega_{63}\sigma_{16}) - \sigma_{13}\Gamma_{13}$$
$$\dot{\sigma}_{14} = i(\Omega_{12}\sigma_{24} + \Omega_{13}\sigma_{34} + \Omega_{14}\rho_{44} + \delta_{14}\sigma_{14} - \Omega_{14}\rho_{11} - \Omega_{24}\sigma_{12} - \Omega_{34}\sigma_{13}) - \sigma_{14}\Gamma_{14}$$
$$\dot{\sigma}_{15} = i(\Omega_{13}\sigma_{35} + \Omega_{15}\rho_{55} + \delta_{15}\sigma_{15} - \Omega_{15}\rho_{11} - \Omega_{35}\sigma_{13}) - \sigma_{15}\Gamma_{15}$$
$$\dot{\sigma}_{16} = i(\Omega_{12}\sigma_{26} + \Omega_{13}\sigma_{36} + \Omega_{16}\rho_{66} + \delta_{16}\sigma_{16} - \Omega_{16}\rho_{11} - \Omega_{26}\sigma_{12} - \Omega_{36}\sigma_{13}) - \sigma_{16}\Gamma_{16}$$
$$\dot{\sigma}_{24} = i(\Omega_{21}\sigma_{14} + \Omega_{24}\rho_{44} + \delta_{24}\sigma_{24} - \Omega_{24}\rho_{22} - \Omega_{14}\sigma_{21}) - \sigma_{24}\Gamma_{24}$$
$$\dot{\sigma}_{26} = i(\Omega_{21}\sigma_{16} + \Omega_{26}\rho_{66} + \delta_{26}\sigma_{26} - \Omega_{26}\rho_{22} - \Omega_{16}\sigma_{21}) - \sigma_{26}\Gamma_{26}$$
$$\dot{\sigma}_{34} = i(\Omega_{31}\sigma_{14} + \Omega_{34}\rho_{44} + \delta_{34}\sigma_{34} - \Omega_{34}\rho_{33} - \Omega_{14}\sigma_{31}) - \sigma_{34}\Gamma_{34}$$
$$\dot{\sigma}_{35} = i(\Omega_{31}\sigma_{15} + \Omega_{35}\rho_{55} + \delta_{35}\sigma_{35} - \Omega_{35}\rho_{33} - \Omega_{15}\sigma_{31}) - \sigma_{35}\Gamma_{35}$$
$$\dot{\sigma}_{36} = i(\Omega_{31}\sigma_{16} + \Omega_{36}\rho_{66} + \delta_{36}\sigma_{36} - \Omega_{36}\rho_{33} - \Omega_{16}\sigma_{31}) - \sigma_{36}\Gamma_{36}$$
(5)

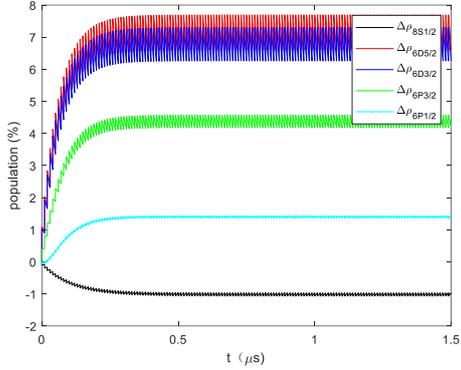

FIG. 6. After considering DTPT, the changes in the population of cesium atoms in the five excited states relative to the DTPT-free process is illustrated.

Here we consider the changes in the population of particles in the excited states after the DTPT process. It can be seen from the Fig. 6 that the population of the excited state including the intermediate state have increased from 1% to about 7%, only the population on 8S$_{1/2}$ state decreases at about 1%. This is because more particles are going from lower states to the three upper states under the direct two-photon reaction with atoms, but atoms at 8S$_{1/2}$ state interact less frequently with others because of its longer lifetime than other states. During the spontaneous decay process, part of the electrons returns to the intermediate state, resulting in an increase in the population of that state. It can be seen from the later analysis that all these processes are possible to be happened. As the total population of particles in the excited state increases at about 17%, the momentum transfer gained by the atoms under the processes of DTPT becomes larger. If this method is used to reduce the movement speed of the atom, its robust effectiveness can be achieved comparing with that under the DTPT-free situation.

## IV. DISCUSSION

Nowadays, the spectrum output from one mode-locked Ti:sapphire laser can easily cover all the above transitions, which ensures DTPT to be happen in multilevel atomic system. The optical frequency of a particular comb mode can be expressed as $f_n = f_{ceo} + nf_{rep}$, where $f_{ceo}$ is the initial frequency, $f_{rep}$ is the pulse repetition rate, and n is the order of the comb which is about several $10^6$ for the transition in Fig.5 for cesium, and similar order for rubidium. For resonant TPTs, atoms are initially at ground state, there are two intermediate P states and three upper states that are highly possible for multiple TPTs to be happen with fixed $f_{ceo}$ while scanning $f_{rep}$. The resonant TPTs have been experimentally demonstrated by Ye's group [9].

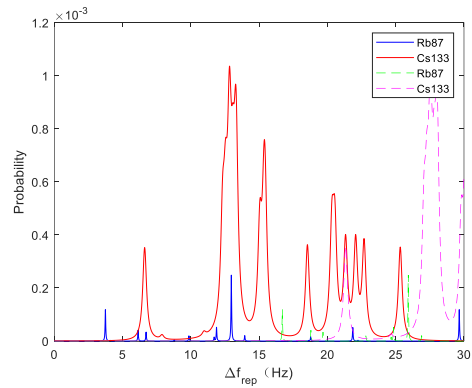

FIG. 7. Calculated transition probability of two alkaline metal atoms when scanning $f_{rep}$, while fixing $f_{ceo}$ of the Mode-locked laser. The dash line indicates the two-photon transition probability spectrum scanned at -0.5frep that the laser frequency is detuned to the resonance transition frequency.

Ye's group used $^{87}$Rb in the experiments, the experimental results are in good agreement with the theoretical results. We now consider that the optical comb acts on two different elements $^{87}$Rb and $^{133}$Cs at the same time. We keep $f_{ceo}$ fixed, and scan $f_{rep}$, one set of transition lines appears every time the range of about 30 Hz is scanned. As shown in Fig.7, it can be clearly distinguished that $^{87}$Rb has 16 fine peaks if changing the vertical axis to logarithm, while only 14 of the 20 spectral lines of $^{133}$Cs can be clearly distinguished. This is because the parameters we choose are for comparison with the experiment of $^{87}$Rb, besides, the frequencies of these transition lines of Cs are very close to each other, resulting in partial mergers. The widening effect can also be seen from the Fig.7.

Theoretically speaking, as long as the total frequency of the two-photon is equal to the frequency between the upper and lower energy levels, a DTPT can occur. It can be seen from Fig.1 that the Rabi frequency is the largest when each photon meets the resonance condition. This is the resonance enhancement of the intermediate energy level. When a two-photon process that does not satisfy resonance enhancement occurs, the transition probability will decrease as the amount of photon frequency detuning from resonance increases. As the dash lines shown in Fig.7, which are the two-photon transition probabilities scanned when the laser frequency is detuned -0.5frep from the resonance transition frequency. when the detuning exceeds one $f_{rep}$, the direct two-photon transition can be ignored.

## V. CONCLUSION

In conclusion, if taking into account the broad spectrum characteristics of the optical FC, the application range of DTPT will be wider. We analyzed the dynamics effects that appear in the three-level atomic system. The Rabi frequency in the DTPT process is proportional to the light intensity, and depends on both the single-photon detuning and the two-photon detuning. The population of particles in the upper energy level presents a comb-like distribution with different speeds. This also leads to the momentum transfer obtained by the atom being related to its moving speed. After considering the DTPT process, the total momentum transfer increase as the number of pulses increase. Considering the transitions of Cs atoms with a wavelength concentrated at 761-921nm, we analyzed the possible DTPT process in the 6-level atomic system and compared it with the case without considering the DTPT process. In general, the population of excited state increases to abut 17%, and the momentum transfer gained by the atom is proportional to the population of particles, which makes a potential cooling effect of the atoms if this technology is used in slowing down the atomic motion with proper configuration. Finally, we analyzed the feasibility of the multi-level system DTPT process. Based on the existing observations, we have obtained a considerable probability of simultaneously inducing Cs and Rb to DTPT. This makes it possible to use optical FCs to simultaneously slow down different elements in the future and provides an alternative solution for ultra-cold molecules. The DTPT process of the optical FCs will also be more widely used in the precision spectroscopy measurement of atoms and cold polar molecules and ultra-cold chemistry in the future.

## ACKNOWLEDGMENTS

This work was supported by the National Natural Science Foundation of China (NSFC) (Grant No. 61535001).

---